# Unveiling the Adsorption and Electronic Interactions of Drugs on 2D Graphsene: Insights from DFT and Machine Learning Approach

Chaithanya Purushottam Bhat, Pranav Suryawanshi, Aditya Guneja, Debashis Bandyopadhyay[*]
Department of Physics, Birla Institute of Technology and Science Pilani
Pilani, Rajasthan - 333031, INDIA
[*]e-mail: Debashis.bandy@gmail.com, bandy@pilani.bits-pilani.ac.in

**Abstract**

Efficient identification of promising drug candidates for nanomaterial-based delivery systems is essential for advancing next-generation therapeutics. In this work, we present a synergistic framework combining density functional theory (DFT) and machine learning (ML) to explore the adsorption behavior and electronic interactions of drugs on a novel 2D graphene allotrope, termed Graphsene (GrS). Graphsene, characterized by its porous ring topology and large surface area, offers an excellent platform for efficient adsorption and strong electronic coupling with drug molecules. A dataset comprising 67 drugs adsorbed on various 2D substrates was employed to train the ML model, which was subsequently applied to predict suitable drug candidates for GrS based on molecular size and adsorption energy criteria (database link provided in a later section). The ML model exhibited robust predictive accuracy, achieving a mean absolute error of 0.075 eV upon DFT validation, though its sensitivity to initialization highlighted the need for larger and more diverse datasets. DFT-based analyses, including adsorption energetics, projected density of states (PDOS), and Bader charge calculations, revealed pronounced charge transfer and electronic coupling between the drug molecules and the GrS surface, elucidating the fundamental nature of drug-substrate interactions. The study reveals that the integrated DFT-ML strategy offers a rapid, cost-efficient approach for screening and understanding drug-nanomaterial interactions, paving the way for data-driven design of advanced nanomaterial-enabled drug delivery systems.

**Graphical Abstract**

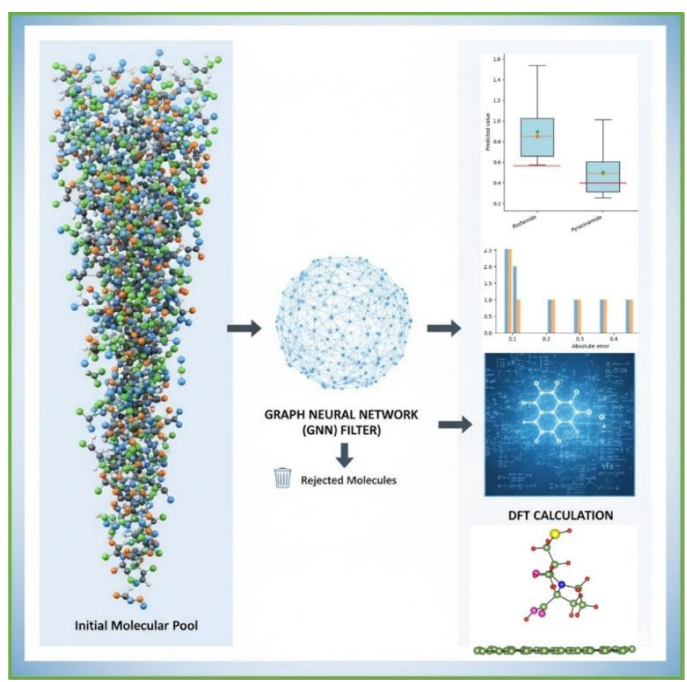

# 1. Introduction:

The development of advanced materials for drug delivery has emerged as a rapidly growing field at the intersection of nanotechnology and biomedicine, providing innovative strategies for precise therapeutic delivery and enhanced clinical performance [1,2]. Conventional drug delivery systems often suffer from limitations such as low bioavailability, poor targeting efficiency, and uncontrolled release rates, which reduce therapeutic effectiveness and increase side effects. In this context, the exploration of two-dimensional (2D) nanomaterials offers a transformative approach due to their unique structural and physicochemical characteristics. Since the discovery of graphene, a surge of research has been directed toward newly synthesized 2D materials with tunable surface properties, exceptional mechanical strength, and chemical versatility [3, 4]. These materials provide ultrathin architectures with large surface-to-volume ratios, which can accommodate a variety of drug molecules through noncovalent adsorption, electrostatic interaction, or chemical functionalization. Their high chemical stability ensures the integrity of the carrier under physiological conditions, while adjustable electronic and surface properties allow for modulation of drug-substrate interactions and controlled drug release profiles. Consequently, the rational design and characterization of novel 2D nanomaterials hold great promise for creating next-generation, high performance drug delivery systems capable of targeted, efficient, and sustained therapeutic action [5-8].

Graphsene (GrS), a newly reported graphene allotrope [9-11], is a carbon-based monolayer composed of fused tetra-, penta-, and dodeca carbon rings. This unique structural arrangement introduces significant porosity and anisotropy, distinguishing GrS from conventional graphene. Such characteristics make it a promising material for drug delivery applications, as the high surface area and tunable surface chemistry enable efficient adsorption of drug molecules and controlled interaction with specific drug functionalities. The use of Density Functional Theory (DFT) provides a powerful approach to evaluate the suitability of GrS for drug delivery prior to experimental studies. Through DFT, one can precisely analyze adsorption energies, electronic structure modifications, and charge transfer processes at the atomic level, offering valuable insights for the rational design of GrS-based drug carrier systems. Nonetheless, large-scale DFT screening across numerous drug candidates remains computationally demanding, which can limit the pace of discovery and optimization in this emerging field [12-14].

To overcome the computational challenges associated with large scale DFT simulations, the integration of machine learning (ML) offers a powerful alternative for the design and optimization of 2D material-based drug delivery systems [15, 16]. A deep learning framework can be developed to efficiently predict adsorption energies, charge transfer characteristics, and interaction mechanisms between 2D materials and various drug molecules [17, 18]. By training these models on high-quality datasets obtained from representative DFT calculations, the ML approach can learn the underlying physical and chemical relationships governing molecular adsorption and electronic behavior. Once trained, such models are capable of rapidly estimating key interaction parameters for a wide range of drug substrate combinations, drastically reducing the computational time required for exhaustive DFT-based evaluations. This enables the efficient screening of large chemical spaces and supports the rational selection of promising drug-material pairs [19-23]. Recent advances in ML, particularly Graph Neural Networks (GNNs), have demonstrated exceptional ability to represent complex atomic structures and capture interatomic interactions with high accuracy. These models can interpret structural descriptors directly from atomic connectivity, making them especially suitable for predicting the adsorption and electronic properties of 2D materials [24-27]. By integrating ML predictions with first-principles DFT validation, the combined framework enhances both the speed and reliability of materials property prediction. This hybrid approach not only accelerates the identification of optimal 2D drug carriers like Graphsene (GrS) but also deepens our understanding of the structure property relationships governing adsorption behavior. Ultimately, the ML assisted design pipeline opens new pathways for data driven discovery and optimization of nanomaterials tailored for next-generation drug delivery applications [28-32].

In the present study, we present an integrated machine learning (ML) and density functional theory (DFT) approach for the accelerated design of efficient two-dimensional (2D) drug delivery systems, focusing on the recently developed graphene allotrope, Graphsene (GrS). The exceptional porosity, anisotropy, and tunable electronic features of GrS make it a promising platform for molecular adsorption and controlled drug release. To harness these properties effectively, ML models were trained on extensive adsorption energy datasets to predict the interaction affinities between a wide range of drug molecules and GrS based substrates. The ML framework employs randomized initialization to ensure model robustness and comprehensive exploration of chemical space, enabling rapid identification of potential drug candidates with favorable binding characteristics. The most promising candidates-selected based on molecular compatibility, optimal adsorption energy, and functional suitability-were subsequently validated through DFT calculations to obtain precise adsorption geometries and energetics. Furthermore, projected density of states (PDOS) and Bader charge analyses were performed to examine the nature of electronic interactions and charge redistribution between the drug molecules and the GrS surface. This combined ML DFT methodology not only reduces the computational cost associated with conventional first-principles simulations but also provides a deeper understanding of the physicochemical mechanisms governing drug adsorption, offering a robust and data driven pathway for the rational design of next-generation 2D nanomaterials for targeted drug delivery applications.

## 2. Methodology:

The objective of this model is to predict the adsorption energy for drug material pairs, which serves as a crucial parameter for assessing their suitability in drug delivery applications. Due to limited data availability, pre-training the model is necessary to achieve robust performance. To further optimize the model, five DFT results using graphene as the substrate were incorporated into the dataset for targeted fine-tuning. This approach leverages the unique properties of graphsene, such as its high surface area and biocompatibility, making it an ideal candidate for drug carrier design. Broadly, the model has two parts a graph-based encoder, and a Multi-Layer Perceptron (MLP) as decoder.

### 2.1. Dataset

In this study, two datasets were employed to facilitate the integrated DFT machine learning analysis of drug adsorption on the novel 2D material, Graphsene. The first dataset, QM9 [33] (Quantum Machine 9), comprising over 100,000 drug-like molecules and 19 physicochemical properties, was utilized to pre-train the encoder network. The second dataset was constructed from previously reported studies on drug material interactions, containing more than 60 pairs of 2D materials and drug molecules along with their corresponding adsorption energies. This combined data framework ensures both broad chemical diversity and specific adsorption information, enabling accurate modeling and prediction of drug Graphsene interactions.

#### 2.1.1. Constructing Molecular Graphs

To represent the chemical and structural information of the systems, all drug molecules were encoded as molecular graphs, where atoms are treated as nodes and chemical bonds as edges. Each node was assigned a comprehensive feature vector that includes a one-hot encoding of the atom type along with key atomic descriptors such as atomic mass, formal charge, electronegativity, and coordination degree, thereby capturing both the elemental identity and topological environment. The molecular graphs were generated using the RDKit [34] library, which ensures efficient and accurate graph construction from molecular structures. It is important to note that these graphs represent the 2D connectivity of the molecules and do not explicitly encode conformational (3D) information. For 2D materials, a slightly modified approach was implemented-atoms were defined as nodes, and edges were established based on a distance cut-off criterion to identify neighboring atoms. Although this method can occasionally introduce additional or missing connections, it does not significantly affect the overall message-passing and learning performance of the graph-based model.

### 2.2. Model

To predict the adsorption energy ($E_d$) of drug molecules on Graphsene, a Graph Convolutional Network (GCN) framework was employed due to its proven capability in effectively learning representations from molecular graph structures. The model architecture consists of two parallel GCN encoders, each dedicated to processing the molecular graph of either the drug or the 2D material. For each graph, two types of pooled feature vectors, maximum pooling and mean pooling, were extracted to capture both the dominant and average structural characteristics. Consequently, four feature vectors (two from each GCN encoder) were obtained and concatenated to form a unified latent representation of the drug material pair. This combined representation was then passed through a Multi-Layer Perceptron (MLP), which performs regression to predict the corresponding adsorption energy. A schematic illustration of the overall model architecture and workflow is provided in Figures 1 and 2.

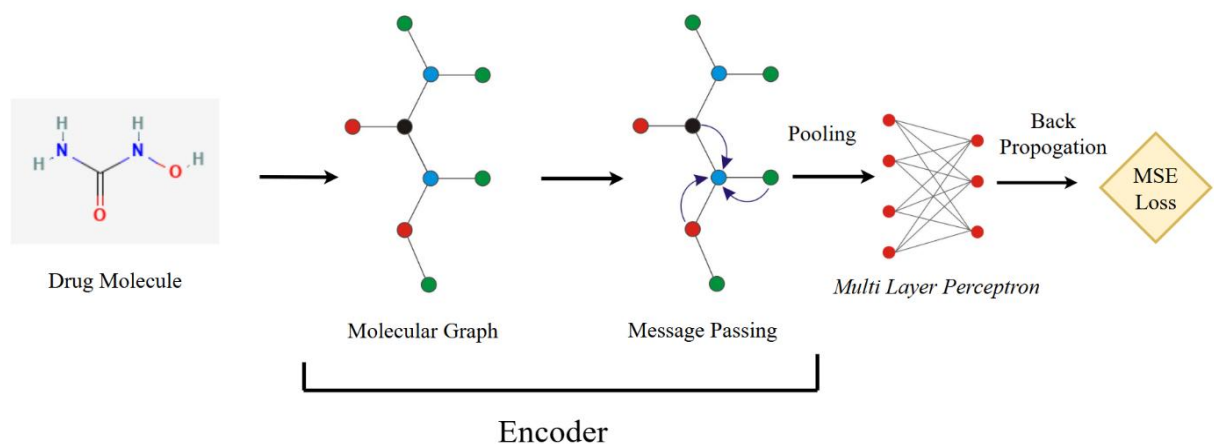

Figure 1. Encoder training pipeline

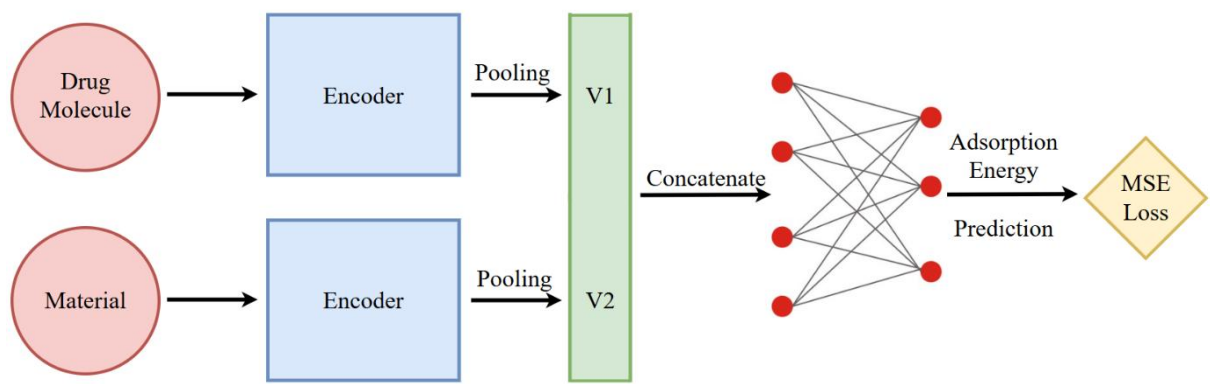

Figure 2. Complete Pipeline

### 2.2.1. Training

The training strategy adopted in this work is based on the transfer learning paradigm, which enhances numerical stability and model generalization, particularly in scenarios with limited data. Initially, a Graph Convolutional

Network (GCN) was trained as an encoder, paired with a Multi-Layer Perceptron (MLP) decoder, using the large scale QM9 dataset. This pre-training phase allowed the encoder to learn rich molecular representations from a diverse set of drug like compounds. After completion of this stage, the pre-trained MLP decoder was discarded, and the encoder was transferred to the new model architecture designed for adsorption energy prediction.

In the second phase, the fine-tuning process was carried out using the drug material adsorption dataset, which contained only 67 data points. In which 5 of the data points contains the results of drug and GrS adsorption energy computed before training the model. These 5 points helps tune the model further. To preserve the learned molecular representations and prevent overfitting, the parameters of the pre-trained encoder were frozen, allowing only the newly added lightweight MLP to be trained on the adsorption data. For both pre-training and fine-tuning phases, the Mean Squared Error (MSE) was employed as the loss function to minimize the deviation between predicted and actual adsorption energies.

**2.3. DFT Validation**

In this study, we have carried out Density Functional Theory (DFT) simulations using the Vienna Ab-Initio Simulation Package (VASP) [35, 36] to study the ground state electronic structure of the Drug-GrS systems. To describe the interaction between ion cores and valence electrons, the projector augmented wave (PAW) method was utilized, which is widely recognized for its accuracy in electronic structure computations [37, 38]. The Perdew-Burke-Ernzerhof (PBE) functional within the generalized gradient approximation (GGA) scheme was employed to account for exchange-correlation interactions [39]. The inclusion of van der Waals interactions is necessary to accurately predict the structural properties. The DFT-D3 method of Grimme with zero-damping function was used to introduce dispersion interactions within the system. The computational parameters are as follows: electronic self-consistency was considered achieved when the energy variation fell below $1.0 \times 10^{-6}$ eV, while ionic relaxation was deemed converged when the force acting on each ion dropped below 0.001 eV/Å. Brillouin zone integration was executed using a $10 \times 10 \times 1$ k-point grid. Furthermore, the plane-wave energy cutoff was set to 520 eV.

**3. Results and discussion:**

**3.1. Structural details of Graphsene (GrS)**

Figure 3 presents the optimized atomic structure and charge density distribution of the Graphsene (GrS) supercell, as obtained from density functional theory (DFT) calculations. The rectangular supercell is composed of forty carbon atoms, arranged within orthogonal lattice dimensions of a = 10.05 Å and b = 12.78 Å. The fully relaxed geometry exhibits excellent structural stability, and the calculated lattice parameters show strong agreement with previously reported values for related carbon based monolayers. This consistency confirms that GrS maintains robust mechanical strength and thermodynamic stability, essential for practical applications.

The GrS monolayer displays a complex polygonal ring architecture, incorporating a combination of tetragonal, pentagonal, and dodecagonal carbon rings. This irregular ring topology introduces significant porosity and anisotropy into the lattice, which distinguishes it from conventional graphene. The open channels and large voids within the GrS framework substantially increase the accessible surface area, while the directional variation in bond orientation leads to anisotropic electronic and mechanical behavior. The charge density analysis further reveals a well distributed electronic cloud throughout the network, indicating strong covalent bonding among carbon atoms and enhanced delocalization of π-electrons, which are favorable for adsorption and charge transfer processes.

This distinctive structural configuration provides several advantages for drug adsorption and delivery applications. The inherent porosity and diverse ring topology create multiple active adsorption sites, enabling strong yet reversible binding of drug molecules through van der Waals, π-π, or electrostatic interactions. Moreover, the high surface to volume ratio of GrS enhances the loading capacity, while its chemical stability and biocompatibility ensure minimal degradation or toxicity under physiological conditions. The combination of these properties makes Graphsene an excellent platform for efficient and controllable drug loading and release. Furthermore, its unique electronic characteristics allow for sensitive detection and monitoring of drugsubstrate interactions, suggesting that GrS could serve as a multifunctional material for next-generation bio-nanotechnological and therapeutic applications.

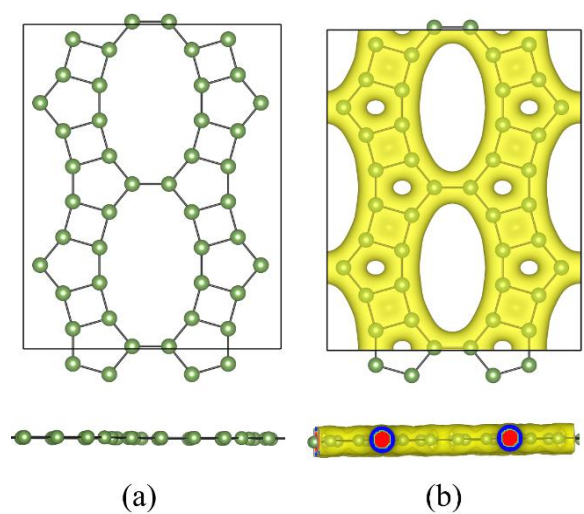

(a)    (b)

Figure 3. Top and side views of (a) the optimized GrS supercell and (b) corresponding charge density distribution.

## 3.2. Convergence of Model

To evaluate the stability and reliability of the machine learning (ML) framework in predicting adsorption energies, a detailed convergence analysis was performed by training and testing the model under multiple random initialization conditions. Initially, a random seed was selected to generate preliminary adsorption energy predictions for a diverse set of drug molecules. Based on these initial predictions, several drug candidates were shortlisted by considering their molecular size (i.e., number of constituent atoms), predicted adsorption energy range, and potential pharmaceutical relevance. Subsequently, density functional theory (DFT) calculations were carried out on these selected candidates to validate the ML predictions, and the comparative results are summarized in Table 1. The accuracy of the ML predictions was quantitatively assessed using the Mean Absolute Error (MAE), defined as:

$$MAE = \frac{1}{N}\sum_{i=1}^{N}\left|E_i^{predicted} - E_i^{DFT}\right| \quad (3.2.1)$$

Where, $E_i^{predicted}$ and $E_i^{DFT}$ represent the predicted and reference (DFT calculated) adsorption energies for the i$^{th}$ drug, respectively, and N is the total number of tested molecules. The computed MAE value of 0.075 eV demonstrates that the ML model exhibits a high degree of accuracy in approximating DFT calculated adsorption energies, effectively identifying potential drug candidates for further theoretical and experimental validation. However, since neural network based models can be sensitive to random weight initialization, the model's performance may vary slightly with different random seeds. To investigate this effect, the model was trained and evaluated using ten distinct random seeds. The resulting distribution of predicted adsorption energies for each seed is illustrated in Figure 4, providing a visual comparison of prediction stability. For each drug, the mean and median predicted values were computed across all seeds, and their deviations from the DFT benchmarks were used to evaluate overall convergence.

The aggregated prediction errors corresponding to each seed are depicted in Figure 5, which reveals that approximately half of the seed configurations produced predictions within 0.1 eV of the DFT reference values. This level of precision confirms that the ML model consistently captures the underlying energy trends of the system. Nevertheless, a few outlier drugs exhibited anomalously high adsorption energies, which can be attributed to either complex adsorption geometries or limited representation of similar chemical environments in the training dataset.

Therefore, the convergence study confirms that the proposed ML model demonstrates robust predictive capability and numerical stability, with reproducible adsorption energy predictions across multiple initializations. This reliable convergence reinforces its applicability as a pre-screening tool for identifying promising drug and Graphsene interaction candidates prior to costly DFT evaluations.

Table1. Comparison of predicted and true energy values for selected drug molecules.

| Drug Name | True Value (eV) | Predicted Energy (eV) | Absolute Error |
|---|---|---|---|
| Ifosfamide ($C_7H_{15}Cl_2N_2O_2P$) | 0.566 | -0.576 | 0.010 |
| Pyrazinamide (Pyrazinoic acid amide) ($C_5H_5N_3O$) | -0.397 | -0.294 | 0.103 |
| Metformin hydrochloride (Glucophage) ($C_4H_{11}N_5 \cdot HCl$) | -0.943 | -1.099 | 0.157 |
| Zidovudine (Retrovir) ($C_{10}H_{13}N_5O_4$) | -0.365 | -0.491 | 0.126 |
| Captopril (Capoten) ($C_9H_{15}NO_3S$) | -0.615 | -0.566 | 0.049 |
| Tiopronin (Thiola) ($C_5H_9NO_3S$) | -0.388 | -0.392 | 0.004 |

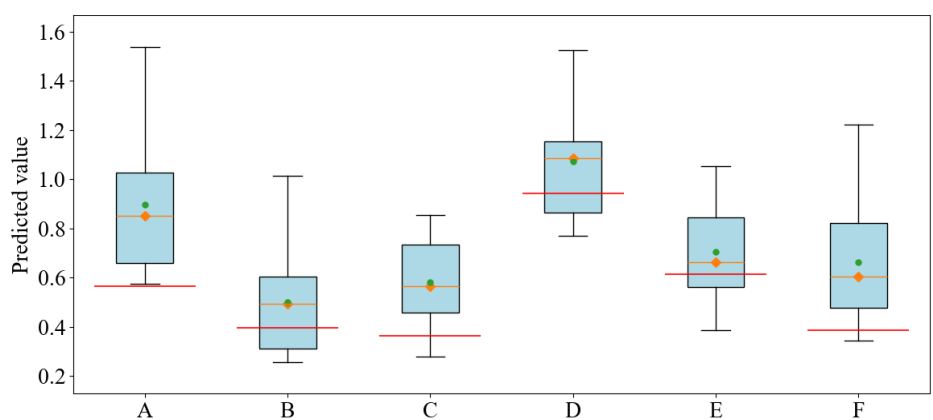

Figure 4. Per-drug prediction distributions across random seeds of (A) GrS+ Ifosfamide, (B) GrS+ Pyrazinamide, (C) GrS+ Zidovudine, (D) GrS+ Metformin hydrochloride, (E) GrS+ Captopril, (F) GrS+ Tiopronin.

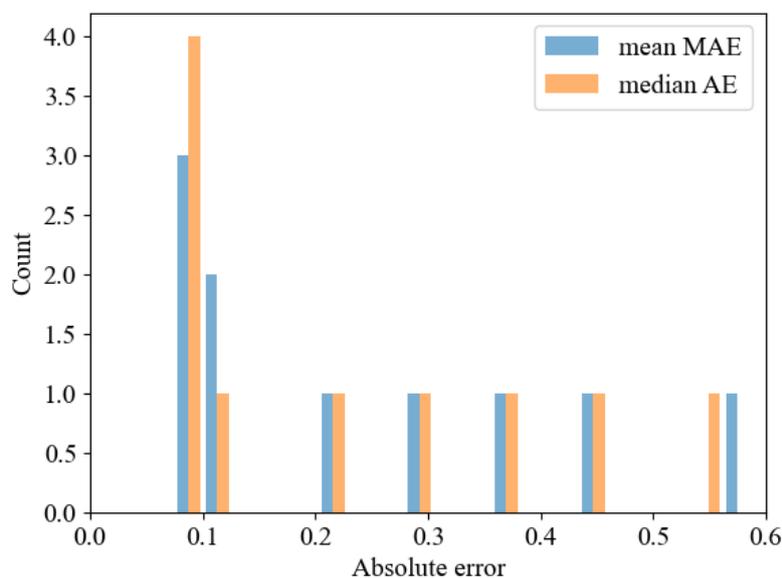

Figure 5. Mean/Median Absolute error for random seeds

### 3.3. Structural Insights into GrS and Drug Interactions

Figure 6 presents the optimized geometries of three representative drug molecules captopril, metformin hydrochloride, and pyrazinamide adsorbed on the Graphsene (GrS) monolayer. The corresponding configurations for Ifosfamide, Zidovudine, and Tiopronin molecules are shown in Supplementary Information Figure 1 (SI Figure 1). To provide a deeper understanding of the electronic response upon adsorption, SI Figure 2 illustrates the charge density distributions of all these drug GrS systems from both top and side perspectives, while SI Figure 3 displays the isolated structures of the individual drug molecules for reference. From the top view, the characteristic pore-like hexagonal framework of the GrS monolayer is distinctly visible, confirming the structural integrity of the substrate after drug adsorption. Each drug molecule exhibits a unique adsorption orientation and interaction pattern depending on its molecular geometry, size, and functional group composition. Captopril, a sulfur- and oxygen-containing molecule, assumes a nearly upright configuration with its heteroatoms (-SH and -COOH groups) oriented toward the GrS surface, indicating potential localized interactions via weak van der Waals forces or hydrogen bonding. In contrast, metformin hydrochloride aligns more parallel to the substrate, suggesting an extended surface contact area that enhances electrostatic and dipole-induced interactions. Pyrazinamide, on the other hand, displays a nearly planar orientation in close proximity to the GrS lattice. This arrangement suggests dominant π-π stacking interactions between the aromatic ring of the drug and the delocalized π-electron network of GrS, leading to a stable adsorption geometry.

The side-view analysis provides further quantitative insights into the adsorption distances and spatial configurations. For captopril, the minimum vertical distance between the molecular backbone and the GrS surface is approximately 2.54 Å, suggesting moderate physisorption accompanied by slight vertical flexibility. Metformin hydrochloride exhibits a slightly larger separation of about 2.66 Å, yet maintains multiple interaction sites across its planar molecular framework. Pyrazinamide, however, shows the most intimate contact with the GrS monolayer, characterized by a minimum adsorption height of 3.38 Å, indicating stronger interfacial coupling primarily driven by π-π interactions. These structural and orientational insights emphasize that the adsorption configuration of each drug molecule on GrS is strongly influenced by its chemical composition and electronic distribution. The variations in adsorption distance, alignment, and orientation directly affect the degree of electronic coupling between the drug and the substrate. Consequently, these parameters are expected to play a crucial role in modulating charge transfer characteristics, adsorption strength, and the potential efficiency of GrS as a drug delivery carrier, as further analyzed in the subsequent sections on electronic structure and charge redistribution.

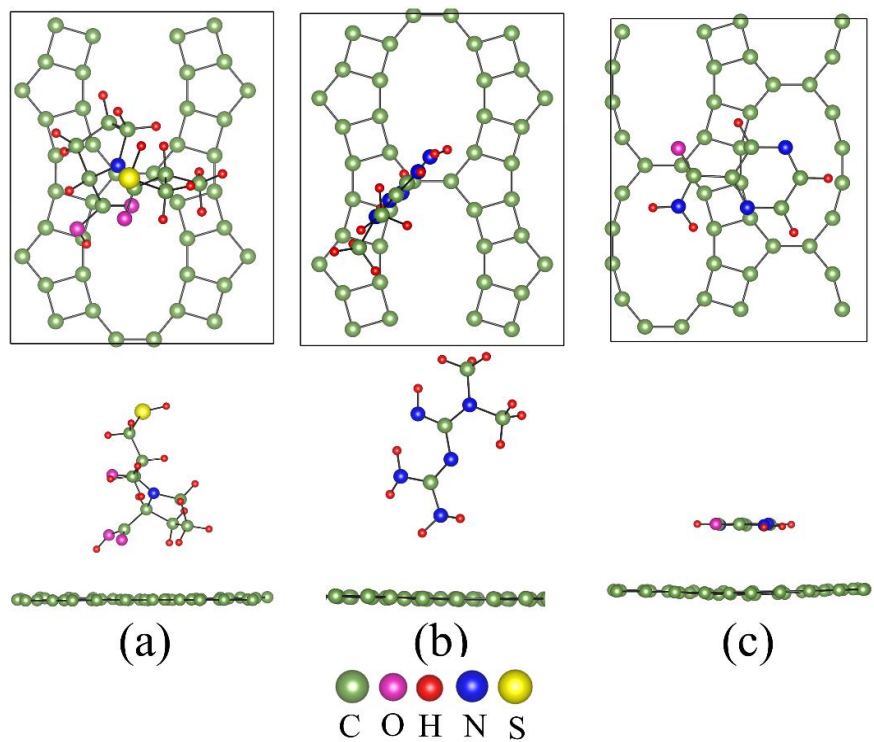

Figure 6. Top and side views of the optimized structures of GrS with adsorbed (a) captopril, (b) metformin hydrochloride, and (c) pyrazinamide.

### 3.4. Partial Density of States analysis

Figure 7 illustrates the partial density of states (PDOS) of pristine GrS and GrS after adsorption of three representative drug molecules: captopril, metformin hydrochloride, and pyrazinamide. The PDOS of pristine GrS exhibits a perfectly symmetric distribution around the Fermi level, with a characteristic vanishing density of states precisely at the Fermi energy ($E_F$). This feature is consistent with the semi-metallic nature of GrS and reflects the presence of Dirac cones at the K-point in its band structure. Such an electronic configuration accounts for the exceptionally high carrier mobility and intrinsic conductivity observed in pristine GrS, confirming its potential as a highly responsive substrate for adsorption-based applications.

After adsorption of captopril on the GrS surface, the PDOS profile undergoes significant modification. New states emerge near the Fermi level, accompanied by peak broadening and asymmetry. These changes arise from orbital hybridization between the molecular orbitals of captopril and the π states of GrS. Moreover, the presence of finite states around $E_F$ indicates charge transfer between the drug molecule and the GrS sheet, effectively altering the local electronic environment. The formation of such hybridized states suggests a strong electronic interaction, which enhances the chemical reactivity of GrS and improves its sensitivity toward captopril adsorption an important feature for sensor-based and drug-delivery applications.

In the case of GrS + metformin hydrochloride, distinct alterations in the PDOS are observed across the energy range from approximately -3 eV to +1 eV relative to the Fermi level. Several new states appear in this range, indicating a moderate degree of electronic coupling and partial charge transfer between the drug and the substrate. The overlap between the electronic states of metformin hydrochloride and those of GrS leads to a redistribution of charge density, implying an effective interaction that modifies the local potential landscape. This behavior suggests that metformin hydrochloride forms a relatively stable adsorption complex on GrS, characterized by both physisorption and weak chemisorption contributions.

For the GrS + pyrazinamide complex, the PDOS exhibits additional, though comparatively less intense, features near the Fermi level. While the changes in PDOS magnitude are less pronounced than those observed for captopril and metformin hydrochloride, the emergence of subtle peaks around $E_F$ points to weak electronic coupling and donor acceptor type interactions between pyrazinamide and the GrS sheet. This mild perturbation reflects a weaker adsorption strength, yet it confirms that even small molecules like pyrazinamide can influence the local electronic structure of GrS through frontier orbital interactions.

So, the PDOS analyses of pure Grs and GrS + Drug systems provide deep insight into the electronic interactions and charge transfer dynamics governing drug adsorption on GrS. The appearance of new electronic states and the redistribution of density near the Fermi level clearly indicate orbital hybridization and charge redistribution between the adsorbates and the substrate. These modifications directly affect the electronic conductivity, surface reactivity, and sensing efficiency of GrS, underscoring its suitability as a multifunctional platform for drug detection and controlled delivery systems.

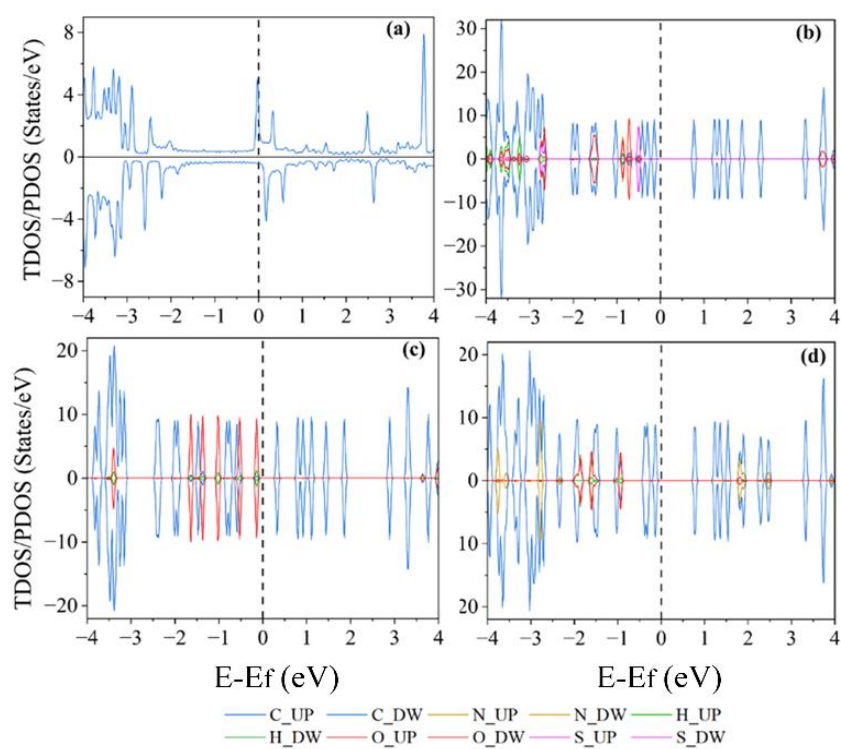

Figure 7. Projected density of states (PDOS) plots for (a) pristine GrS, (b) GrS+Captopril, (c) GrS+Metformin hydrochloride and (d) GrS+Pyrazinamide.

### 3.5. Bader charge analysis:

To further elucidate the electronic interactions between the adsorbed drug molecules and the substrate, Bader charge analysis [40] was performed. The charge transfer upon adsorption ($\Delta q$) quantifies the redistribution of electron density between the drug and substrate. Here, a negative $\Delta q$ value indicates that the drug acts as an electron acceptor (gaining electrons), while a positive value implies electron donation to the substrate. The Bader charge differences calculated for the investigated drug molecules are summarized in Table 2, while the comprehensive dataset is provided in SI Table 1 and 2.

The modifications in the PDOS following drug adsorption are in excellent agreement with the trends observed in the Bader charge analysis. For instance, the substantial emergence of new states near the Fermi level and pronounced peak broadening in the captopril-adsorbed system correlate with the largest negative Bader charge value $\Delta q = -0.066$, confirming significant electron transfer from graphene to the drug. Similarly, the moderate PDOS changes seen for metformin hydrochloride are consistent with its mild electron-donor behavior $\Delta q = 0.051$, while the relatively subtle PDOS alterations in the pyrazinamide system reflect the minimal charge transfer observed $\Delta q = -0.009$. Together, these results demonstrate that the degree of charge transfer, as quantified by Bader analysis, directly influences the extent of electronic structure modification in graphene upon drug binding. This interplay between orbital hybridization and charge redistribution underpins both the strength and character of the adsorption process, ultimately impacting the material's efficacy in sensing or drug-delivery applications.

**Table 2.** Calculated Bader charge difference (or charge exchange) ($\Delta q$ in e) between GrS and the adsorbed drug.

| Systems | $\Delta q$ in e |
|---|---|
| GrS+Captopril | -0.066 |
| GrS+Metformin hydrochloride | 0.051 |
| GrS+Pyrazinamide | -0.009 |

### 4. Conclusion

This study establishes a comprehensive and data-driven computational framework for evaluating drug adsorption on Graphsene (GrS) monolayers by integrating graph neural network (GNN) based machine learning (ML) predictions with density functional theory (DFT) validation. The developed workflow efficiently combines predictive screening and first-principles confirmation, thereby offering an accelerated route for identifying promising drug substrate interactions in complex chemical spaces.

Through systematic sampling of random seed initializations, the ML model demonstrated a high degree of predictive capability, accurately estimating adsorption energies across diverse drug molecules. The model achieved an impressive mean absolute error (MAE) of 0.075 eV, underscoring its reliability in capturing the essential physicochemical trends governing adsorption. However, the results also revealed minor sensitivity to random initialization and outlier responses for certain molecular systems, suggesting that further refinement is needed. Expanding the training dataset to include a more chemically diverse library, improving molecular descriptors, and fine-tuning the network architecture would enhance the robustness and generalization of the ML framework across broader chemical domains.

Complementary, DFT investigations, including projected density of states (PDOS) and Bader charge analyses, provided in-depth mechanistic insights into the adsorption phenomena. The PDOS profiles revealed the emergence of new hybridized electronic states and a redistribution of charge density near the Fermi level upon drug adsorption, signifying strong orbital hybridization and charge transfer between the adsorbates and the GrS surface. Among the examined drug molecules, captopril exhibited the most pronounced electronic coupling, followed by metformin hydrochloride and pyrazinamide, indicating variable interaction strengths governed by their electronic structure and functional groups. These findings elucidate the fundamental electronic origins of the binding behavior and highlight how molecular properties modulate GrS's chemical reactivity and sensor responsiveness.

The integrated ML DFT framework demonstrates a powerful approach for the rapid, accurate, and interpretable screening of drug adsorption on 2D materials. The synergy between data-driven predictions and quantum mechanical validation establishes a scalable methodology for rational design and optimization of nanomaterial drug interfaces. Looking forward, extending this framework with more diverse chemical datasets, improved structural representations, and experimental corroboration will further enhance predictive accuracy and practical applicability. Together, the present work underscores the potential of combining machine learning with quantum-level modeling to accelerate the discovery of next generation nanomaterials for biomedical applications, including drug delivery, biosensing, and molecular recognition. The insights gained from this study not only advance our understanding of drug-GrS interactions but also lay the groundwork for future exploration of AI-guided materials design in emerging nano-bio interfaces.

**Acknowledgments**

The authors gratefully acknowledge the Anusandhan National Research Foundation (ANRF), Government of India, for the research support provided through the grant CRG/2022/003249, which made this work possible.

**CRediT author statement**

Chaithanya Purushottam Bhat contributed primarily to VASP simulations, data analysis, figure preparation, reference management, and manuscript review and revision. Pranav Suryawanshi writing draft, training ML model, editing graphical. Aditya Guneja literature survey, preparation of dataset. Debashis Bandyopadhyay, the corresponding author, was responsible for overall project coordination, conceptual design, VASP computations, data interpretation, manuscript drafting and editing, and project administration. These CRediT roles provide a clear and transparent account of each author's contributions to the research.

**Appendix A. Supplementary data**

Supplementary data to this article can be found online.

**Declaration of interests**

The authors confirm that they have no financial interests or personal relationships that could be perceived as conflicting with the research presented in this paper.

**Declaration of Generative AI and AI-assisted technologies in the writing process:**

To enhance language quality and ensure proper citation, the authors utilized Grammarly and Turnitin during the manuscript preparation. Following the use of these tools, the authors carefully reviewed and revised the content as needed and take full responsibility for the accuracy and integrity of the final manuscript.

**Data availability declaration**

All data are provided in the manuscript, presented both in tables and graphical format.